\documentclass[a4paper,fleqn,usenatbib]{mnras}

\usepackage[T1]{fontenc}
\usepackage{ae,aecompl}
\usepackage{graphics,epsf}
\usepackage{amsmath}                
\usepackage{amsfonts}               
\usepackage{amssymb}                
\usepackage{epsfig}
\usepackage{epstopdf}
\usepackage{multirow}
\usepackage[para,online,flushleft]{threeparttable}
\usepackage{flushend}

\title[Leftover hydrogen in stripped massive stars]{Effects of winds on the leftover hydrogen in massive stars following Roche lobe overflow}
\author[Gilkis, Vink, Eldridge \& Tout]{Avishai Gilkis$^{1}$\thanks{E-mail: \href{agilkis@ast.cam.ac.uk}{agilkis@ast.cam.ac.uk} (AG); \href{jorick.vink@armagh.ac.uk}{jorick.vink@armagh.ac.uk} (JSV);
\href{j.eldridge@auckland.ac.nz}{j.eldridge@auckland.ac.nz} (JJE)}, Jorick S. Vink$^{2}$, J.~J. Eldridge$^{3}$, Christopher A. Tout$^{1}$
\\
$^{1}$ Institute of Astronomy, University of Cambridge, Madingley Road, Cambridge, CB3 0HA, UK\\
$^{2}$ Armagh Observatory and Planetarium, College Hill, Armagh, BT61 9DG, UK\\
$^{3}$ Department of Physics, University of Auckland, Private Bag 92019, Auckland 1010, New Zealand
}

\date{Accepted 2019 April 18. Received 2019 April 18; in original form 2018 November 5}

\pubyear{2019}

\volume{486}

\begin{document}

\pagerange{4451--4462}

\maketitle
\label{firstpage}

\begin{abstract}
We find that applying a theoretical wind mass-loss rate from Monte Carlo radiative transfer models for hydrogen-deficient stars results in significantly more leftover hydrogen following stable mass transfer through Roche lobe overflow than when we use an extrapolation of an empirical fit for Galactic Wolf-Rayet stars, for which a negligible amount of hydrogen remains in a large set of binary stellar evolution computations. These findings have implications for modelling progenitors of Type~Ib and Type~IIb supernovae. Most importantly, our study stresses the sensitivity of the stellar evolution models to the assumed mass-loss rates and the need to develop a better theoretical understanding of stellar winds.
\end{abstract}


\section{INTRODUCTION}
\label{sec:intro}

A large fraction of massive stars, if not all of them, end their lives as energetic and luminous supernovae (SNe). These core-collapse SNe (CCSNe) are classified according to their observed light curves and spectral features. One of the most conspicuous characteristics for CCSN classification is the presence of hydrogen spectral lines \citep{Filippenko1997}. Type~II SNe are those for which hydrogen is observed, while hydrogen is absent in spectra of Type~I SNe (specifically Ib and Ic, for massive stars). A special case of Type~II SNe, for which hydrogen spectral features are observed at early stages but later disappear, is termed IIb (e.g. SN~1993J, \citealt{Nomotoetal1993}).

\cite{Dessartetal2011} find, in their modelling, that a total hydrogen mass in the stellar envelope of  $M_\mathrm{H} \ga 0.001 \, \mathrm{M}_\odot$ results in a Type~IIb SN rather than a Type~Ib. Type~Ib and Type~IIb CCSNe are thought to arise from similar evolutionary channels which result in a small amount of hydrogen left in the stellar envelope \citep*{Yoonetal2017}. Recent studies give priority to binary evolution channels for progenitors of both Type~Ib \citep{Yoon2015} and Type~IIb (e.g. \citealt{Podsietal1993,Claeysetal2011,Soker2017}), though single-star progenitors are not ruled out \citep{KotakVink2006,Yoonetal2012}. It is noteworthy that the remnant Cassiopeia A, which is agreed to have been a Type~IIb, contains no remaining companion star \citep{Kochanek2018,Kerzendorfetal2019}.

The basic scenario for Type~Ib and Type~IIb SNe in binary systems \citep{Yoonetal2017} is that the primary star expands as it evolves until it fills its Roche lobe, when mass transfer starts owing to Roche lobe overflow (RLOF). If the mass transfer is stable it continues until a small amount of hydrogen is left in the envelope of the primary, at which point the star starts to shrink. Further mass loss is through stellar winds\footnote{Additional RLOF episodes can occur, for certain initial parameters, when the primary expands again.}. The amount of hydrogen left, if any, depends on the assumed mass-loss rate at this stage. Here we aim to emphasize the importance of the post-RLOF stellar winds, and the associated uncertainties, for the properties of stellar model envelopes and the leftover hydrogen in them.

\section{METHOD}
\label{sec:method}

We use the Modules for Experiments in Stellar Astrophysics code (\textsc{mesa}, version 10398, \citealt{Paxton2011,Paxton2013,Paxton2015,Paxton2018}) to evolve binary stars with a metallicity of $Z=0.019$ from the main sequence until the end of carbon burning in the core of the primary. This stage is just years before iron core collapse and the properties of the outer parts of the star are not expected to change \citep*{Woosleyetal2002}. We ran models with initial primary masses of $M_1 / \mathrm{M}_\odot \in \{12,14,16,19,22,25\}$, secondary masses listed in Table \ref{tab:masses} and orbital periods of $P_\mathrm{i} / \mathrm{d} \in \{5,10,18,33,60,110,201,367,669,1219,2223\}$.
\begin{table}
\centering
\caption{Initial masses for stellar evolution calculations}
\begin{threeparttable}
\begin{tabular}{c|ccc}
\hline
& $0.9>q>0.8$ & $0.7>q>0.6$ & $0.45>q>0.35$ \\
$M_1/ \mathrm{M}_\odot$ & $M_2/ \mathrm{M}_\odot$ & $M_2/ \mathrm{M}_\odot$ & $M_2/ \mathrm{M}_\odot$ \\
\hline
$12$ & $10$  & $8$  & $5$ \\
$14$ & $12$  & $9$  & $5$ \\
$16$ & $14$  & $10$ & $6$ \\
$19$ & $16$  & $12$ & $7$ \\
$22$ & $19$  & $14$ & $8$ \\
$25$ & $22$  & $16$ & $9$ \\
\hline
\hline
\end{tabular}
\footnotesize
\label{Table:Runs}
\end{threeparttable}
\label{tab:masses}
\end{table}
Effects related to rotation, such as rotational mixing and tidal synchronization, are not taken into account. This choice was made to concentrate on the effect under study in a simple manner. The following sub-sections detail the main aspects of the stellar modelling and additional technical details on code implementations are in Appendix~\ref{app:code}.

\subsection{Wind mass loss}
\label{subsec:winds}

Mass loss through winds is according to \cite{Vink2001} for hydrogen-rich hot stars and according to \cite{deJager1988} for effective surface temperatures below $10^4 \, \mathrm{K}$. For hot stars with hydrogen surface abundances $X_\mathrm{s}$ below $0.4$ we use one of two different mass-loss prescriptions, either that of \cite{V17} or \cite{NL00}. The mass-loss rate of \cite{V17}, which we herein refer to as V17, is
\begin{eqnarray}
\begin{aligned}
\log_{10} \left( \dot{M} /  \mathrm{M}_\odot \, \mathrm{yr}^{-1} \right) = -13.3 + 1.36 \log_{10} \left( L /  \mathrm{L}_\odot \right) \\ + \, 0.61 \log_{10} \left( Z_\mathrm{s} / 0.019 \right),
\label{eq:MdotV17}
\end{aligned}
\end{eqnarray}
where $\dot{M}$ is the mass-loss rate, $L$ is the stellar luminosity and $Z_\mathrm{s}$ is the metallicity at the photosphere. The mass-loss prescription following \cite{NL00}, herein NL00, is
\begin{eqnarray}
\begin{aligned}
\log_{10} \left( \dot{M} /  \mathrm{M}_\odot \, \mathrm{yr}^{-1}\right) =  -11.0 + 1.29 \log_{10} \left( L /  \mathrm{L}_\odot \right) \\ + \, 1.7 \log_{10} \left( Y_\mathrm{s} \right) + 0.5 \log_{10} \left( Z_\mathrm{s} \right),
\label{eq:MdotNL00}
\end{aligned}
\end{eqnarray}
where $Y_\mathrm{s}$ is the surface helium abundance. We note that the V17 prescription has no dependence on $Y_\mathrm{s}$ because this is considered to be unrelated to the physics of the wind driving.

The NL00 recipe is based on empirical modelling of observed Wolf-Rayet (WR) stars. Unfortunately only a few stripped stars are known so we cannot rely on empirical rates for lower-mass stripped helium stars as we can for classical WR stars\footnote{The NL00 prescription has some difficulties with classical WR stars as well \citep{Yoon2017}.}. The one exception could be HD45166 \citep{qWR} but this system might have undergone a different evolution from the simple RLOF we model here. There are too few actual measured mass-loss rates for stripped stars \citep{Yoon2015} to derive a reliable empirical mass-loss rate prescription for helium stars with masses and luminosities lower than those of the classical WR stars. One option to overcome this observational inadequacy is to extrapolate the NL00 recipe towards the regime of lower masses and lower luminosities but the dependencies on helium abundance $Y_\mathrm{s}$ and total metallicity $Z_\mathrm{s}$ in NL00 are thought to be unphysical (e.g. \citealt{PulsVinkNajarro2008}). Extrapolation of the NL00 recipe to a parameter regime for which it was not derived is then of rather limited value.

\cite{V17} makes a pilot study that provides theoretical predictions for stripped helium stars using Monte Carlo models with a fixed effective temperature of $50\,000\, \mathrm{K}$. These winds remain optically thin in the simulated parameter range and it remains to be seen whether the winds become optically thicker at higher effective temperatures. A higher effective temperature $T_\mathrm{eff}$ would imply a smaller star which is more likely to become optically thick so that the mass-loss rate might increase substantially. If the winds were to remain optically thin at higher $T_\mathrm{eff}$ we would not expect the mass-loss rate to change dramatically unless there is insufficient line opacity at higher $T_\mathrm{eff}$ or there is an opacity or bi-stability jump, as found for hydrogen-rich stars at lower $T_\mathrm{eff}$ \citep{Vink1999}.

In any case a transition between optically thin stripped helium stars and optically thick WR stars might be expected somewhere in the helium star regime, similar to the mass-loss kink in the hydrogen-rich part of the Hertzsprung--Russell diagram \citep{Vinketal2011, Bestenlehneretal2014}. Although more work is needed to cover the entire parameter space and to investigate and scrutinize the accuracy of the \cite{V17} pilot study, we consider the order-of-magnitude lower mass-loss rates provided by this theoretical scheme compared to the simple extrapolations of NL00 to lower masses and luminosities of stripped stars to be more likely to be correct.

\subsection{Mass transfer by Roche lobe overflow}
\label{subsec:rlof}

The mass-transfer rate by RLOF $\dot{M}_\mathrm{tr}$ is calculated according to the scheme of \cite{KolbRitter1990}. We implement an updated mass-transfer scheme\footnote{See Appendix \ref{app:code} for details.} so that the mass-transfer efficiency is limited by the thermal time-scale of the accretor,
\begin{eqnarray}
\begin{aligned}
\beta \dot{M}_\mathrm{tr} \le M / \tau_\mathrm{th},
\label{eq:Mdotth}
\end{aligned}
\end{eqnarray}
where $\beta$ is the mass transfer efficiency and the thermal time-scale is defined by
\begin{eqnarray}
\begin{aligned}
\tau_\mathrm{th} \equiv \frac{G M^2}{L R},
\label{eq:tth}
\end{aligned}
\end{eqnarray}
with $R$ and $M$ the photospheric radius and mass of the primary. In addition our mass transfer efficiency smoothly drops to zero if the radius of the secondary enters the range $0.99 < R_2 /R_{\mathrm{L},2} < 1.0$, where $R_2$ is the photospheric radius of the secondary and $R_{\mathrm{L},2}$ is its Roche lobe radius\footnote{This is usually avoided by following equation (\ref{eq:Mdotth}) but not in all cases. These limitations on the mass accretion arise because the material is not tightly bound by the gravity of the secondary and so is assumed to be lost from the system.}. Otherwise the mass transfer efficiency is $0.9$, though in theory it might also be reduced owing to the spin-up of the secondary \citep{Packet1981}. Tidal synchronization also affects the orbital evolution in close systems. Our modelling assumptions were chosen to allow for a large range of initial and final conditions to be investigated. For the current purpose, of demonstrating the effect of stellar winds, this is sufficient.  

\subsection{Orbital angular momentum}
\label{subsec:jorb}

The orbital separation evolves as angular momentum is lost from the system, affecting the widening of the system and so the occurrence of late mass-transfer episodes. Material lost from the primary in a wind carries away the specific angular momentum of the orbit of the primary. The stellar wind of the secondary also leads to angular momentum loss but less so. Material lost from the system because of inefficient mass transfer, as described in Section \ref{subsec:rlof} and Appendix \ref{app:code}, carries away the specific angular momentum of the secondary.

\subsection{Mixing}
\label{subsec:mix}

The Ledoux criterion is applied to define convective regions, in which mixing is according to a mixing-length theory \citep{MLT}, with $\alpha_\mathrm{MLT}=1.5$. Semiconvection is according to \cite{Langer1983}, with an efficiency parameter of $\alpha_\mathrm{sc}=1.0$. Overshooting above convective regions is as by \cite{Herwig2000}. We include thermohaline mixing by the method of \cite{thermohaline}.

The dependence of our results on the initial parameters quantitatively changes for different assumptions for the mixing processes. \cite{Sukhboldetal2018} find that the time a stellar model spends as a blue super giant relative to the time it spends as a red super giant depends on semiconvection so we might expect that, with less efficient semiconvective mixing, stellar models would reach large radii at earlier times, changing the dependence of our results on the initial periods. Augmented overshooting increases the helium core masses. Rotational mixing, which we do not account for, can similarly affect our results. The various mixing processes, as well as the definitions of convective boundaries, can affect the stellar mass-loss rate through the composition dependence in equation (\ref{eq:MdotNL00}).

\section{RESULTS}
\label{sec:results}

\begin{figure*}
\includegraphics[width=1.0\textwidth]{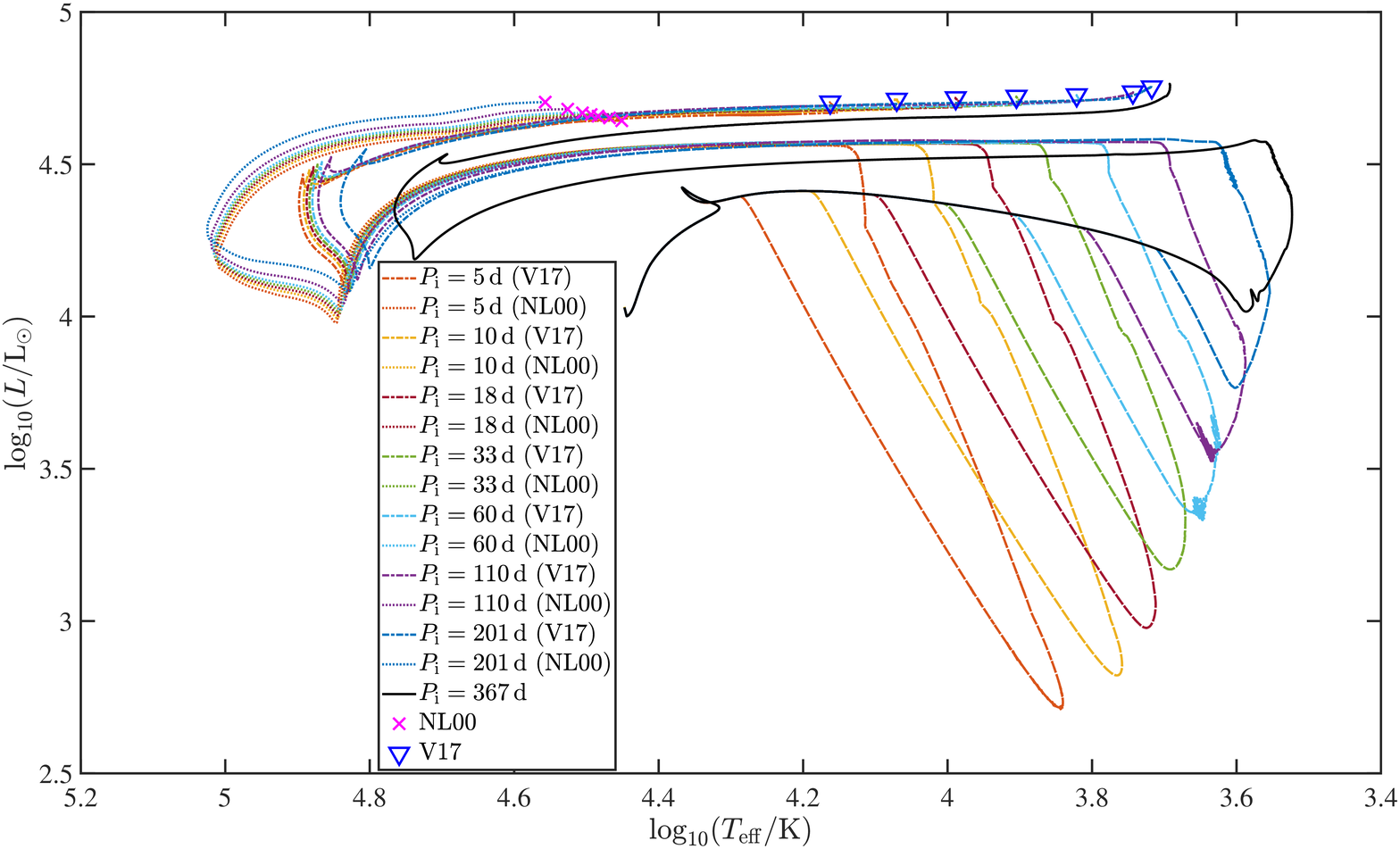}
\caption{Evolutionary tracks on the Hertzsprung-Russell diagram for models with a primary mass of $M_1 = 12 \,  \mathrm{M}_\odot$ and a secondary initial mass of $M_2 = 10 \,  \mathrm{M}_\odot$ with the initial orbital periods and mass-loss prescriptions indicated in the inset. For initial periods of $201 \, \mathrm{d}$ and shorter two different mass-loss rates were used after the surface hydrogen abundance dropped below $0.4$. The evolutionary end points are marked with triangles for the V17 models and crosses for the NL00 models.}
\label{fig:HR1}
\end{figure*}
Fig. \ref{fig:HR1} shows several example evolutionary tracks on a Hertzsprung--Russell diagram for an initial primary mass of $M_1 = 12 \,  \mathrm{M}_\odot$ and a secondary initial mass of $M_2 = 10 \,  \mathrm{M}_\odot$, several initial orbital periods between $5$ and $367 \, \mathrm{d}$ and the different mass-loss schemes for hydrogen-deficient stars discussed in Section~\ref{subsec:winds}. For initial periods of $P_\mathrm{i} \ge 367 \, \mathrm{d}$ the surface hydrogen abundance remained above $0.4$ throughout the evolution. For $P_\mathrm{i} < 367 \, \mathrm{d}$ the evolutionary tracks diverge after the surface hydrogen abundance drops below $0.4$: models with the NL00 mass-loss rate end with significantly higher effective surface temperatures. The V17 models are cooler and have larger photospheric radii owing to a small but non-negligible amount of hydrogen left in their envelopes, as discussed below. The V17 models with $P_\mathrm{i} \le 33 \, \mathrm{d}$ in Fig. \ref{fig:HR1} experience a second phase of mass transfer after filling their Roche lobes as helium giants. The V17 models with $60 \le P_\mathrm{i}/ \mathrm{d} \le 201$ in Fig. \ref{fig:HR1} are close to filling their Roche lobes. Whether a second phase of mass transfer commences depends on the initial masses and can occur also for initial periods longer than $60\,\mathrm{d}$ (see Appendix \ref{app:models}). Hereinafter all models discussed and presented are those for which the surface hydrogen abundance drops below $0.4$ during the evolution so that the NL00 and V17 mass-loss prescriptions are switched on. Models with similar characteristics might also result when $X_\mathrm{s}>0.4$.

\begin{figure*}
\includegraphics[width=1.0\textwidth]{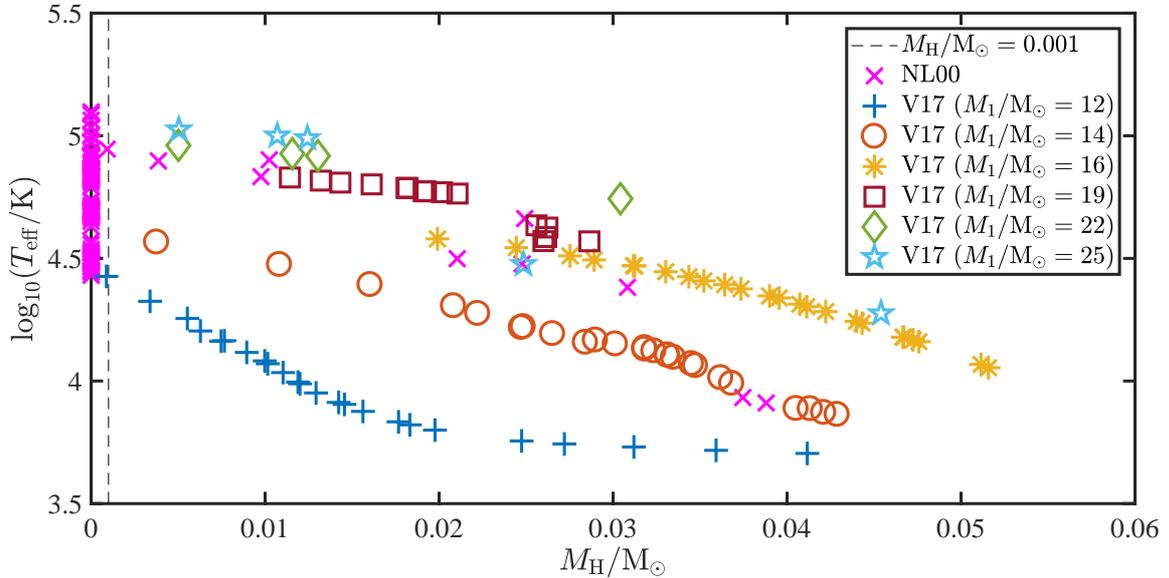}
\caption{Effective surface temperature $T_\mathrm{eff}$ as function of total hydrogen mass $M_\mathrm{H}$ in the final stellar models for which the surface hydrogen abundance is less than $0.4$. The initial mass of the primary is indicated for the V17 models. The vertical dashed line indicates the threshold of $M_\mathrm{H} \ga 0.001 \, \mathrm{M}_\odot$ for Type~IIb SNe \citep{Dessartetal2011}.}
\label{fig:TvsMH}
\end{figure*}
Fig. \ref{fig:TvsMH} shows the final effective surface temperature for all the models which reached carbon depletion with $X_\mathrm{s} < 0.4$ as a function of the total leftover hydrogen mass $M_\mathrm{H}$. Almost all models with the NL00 mass-loss rate ended up with virtually no hydrogen left at all, while the models for which the V17 mass-loss rate was used all have $M_\mathrm{H}>0.0008 \, \mathrm{M}_\odot$ and most have $M_\mathrm{H}>0.01 \, \mathrm{M}_\odot$ at the end. This is because the V17 prescription results in post-RLOF mass-loss rates about an order of magnitude lower than the NL00 prescription.

There are several apparent trends in Fig. \ref{fig:TvsMH}. The sequence for lower temperatures is for models with final masses in the range $2.95 < M / \mathrm{M}_\odot < 3.41$ (helium core masses of $2.92 < M_\mathrm{c} / \mathrm{M}_\odot < 3.25$), the mid-temperature sequence is for models with final masses in the range $3.75 < M / \mathrm{M}_\odot < 4.3$ (helium core masses of $3.69 < M_\mathrm{c} / \mathrm{M}_\odot < 4.12$) and the sequence at the top is for the higher mass models with $4.74 < M / \mathrm{M}_\odot$ (helium core masses of $4.6 < M_\mathrm{c} / \mathrm{M}_\odot$). This is because more massive helium cores are hotter, while an extended envelope above them reduces the effective photospheric temperature.

\begin{figure*}
\includegraphics[width=1.0\textwidth]{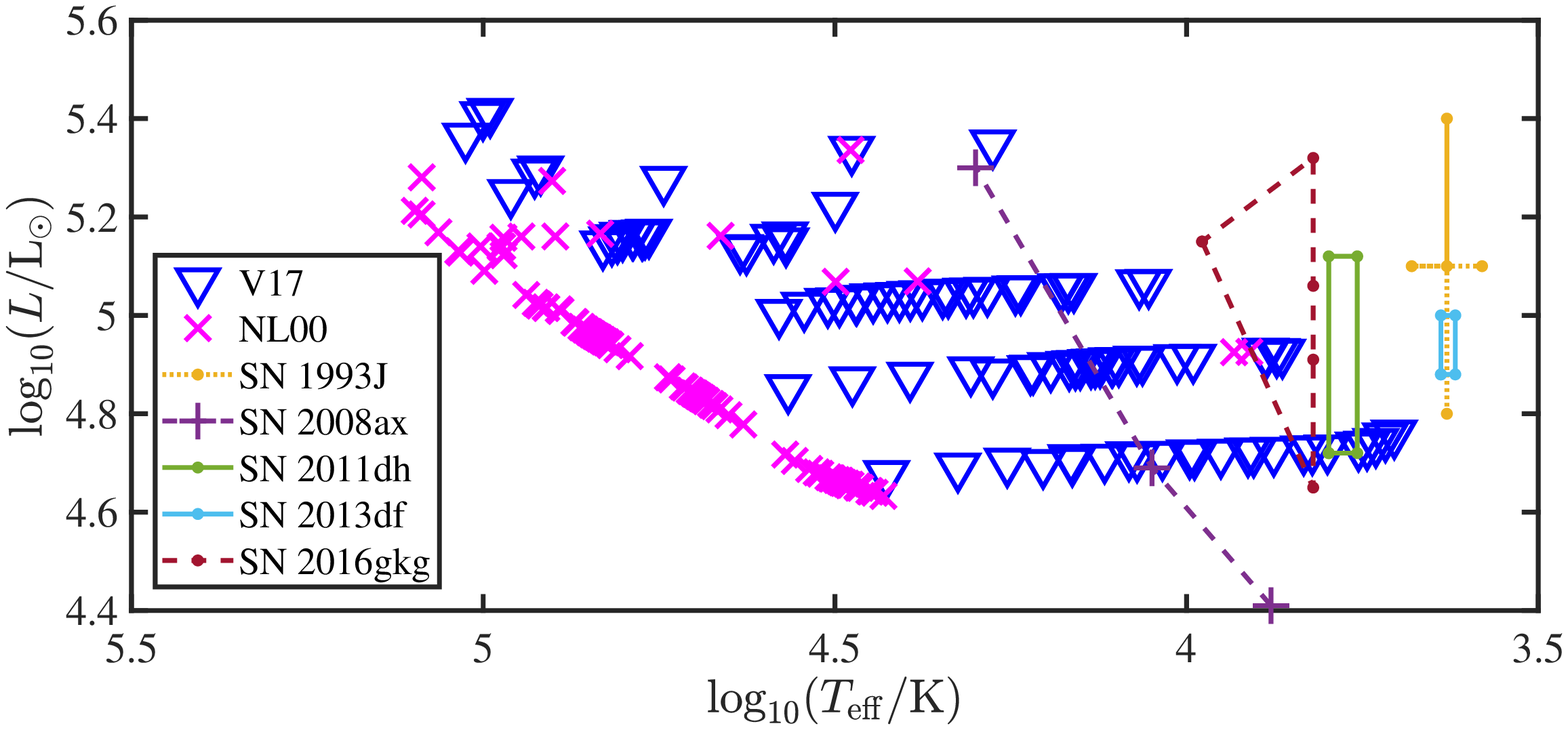}
\caption{Luminosity as function of effective surface temperature for all models with $X_\mathrm{s} < 0.4$. Also plotted are progenitors of Type~IIb SNe, SN~1993J \citep{Maundetal2004}, SN~2008ax \citep{Folatellietal2015}, SN~2011dh \citep{Maundetal2011}, SN~2013df \citep{VanDyketal2014}, and SN~2016gkg (\citealt{Kilpatricketal2017} for the hotter estimate; \citealt{Tartagliaetal2017} for the cooler points; see also \citealt{Arcavietal2017}).}
\label{fig:HR2}
\end{figure*}
Fig. \ref{fig:HR2} shows how the luminosity of the final models varies with their effective surface temperature. The NL00 models mostly follow a well-defined sequence, similar to that reported by others (e.g., \citealt{Yoonetal2017}). The V17 models group into several sequences according to their mass. Observational properties of Type~IIb progenitors are plotted, with the hotter falling near our V17 models while the cooler probably have slightly more hydrogen in their envelopes.

\begin{figure}
\includegraphics[width=0.47\textwidth]{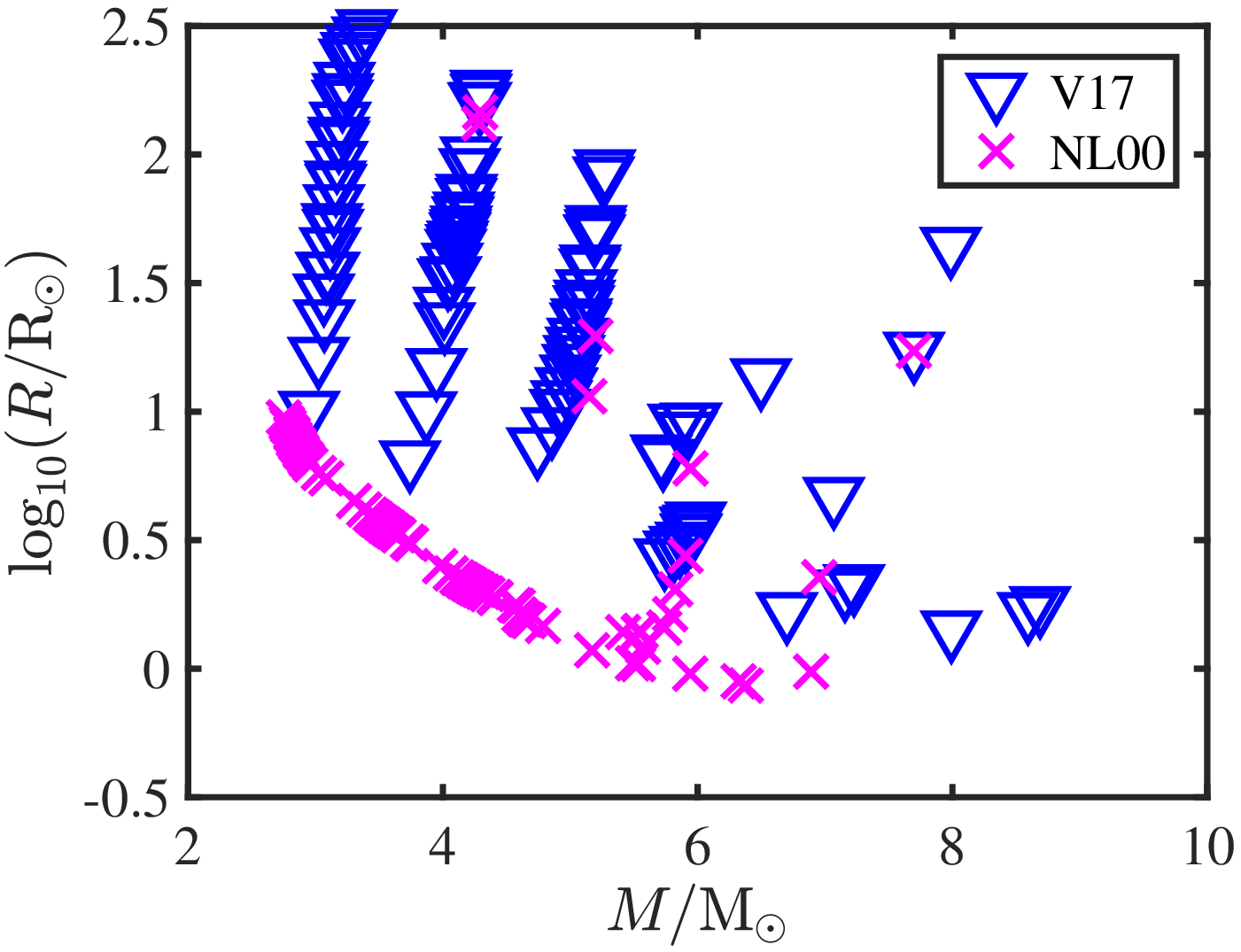}
\caption{Photospheric radius as function of final mass for all models with $X_\mathrm{s} < 0.4$.}
\label{fig:RvsM}
\end{figure}
Fig. \ref{fig:RvsM} shows the photospheric radii of the final models as a function of the final stellar mass. The NL00 models mostly follow an inverse mass--radius relation, as has been reported for evolved helium stars (e.g., \citealt{Habets1986,Yoonetal2017}). The V17 models tend to become much more expanded because of their hydrogen envelopes. The sequences of V17 models with a narrow mass range and large range in radii correspond to the trends seen in Fig. \ref{fig:TvsMH} and discussed above. All models follow very closely the same mass--luminosity relation, $\log_{10} \left( L /  \mathrm{L}_\odot \right) \simeq 1.6 \log_{10} \left( M /  \mathrm{M}_\odot\right) + 3.9$ for $3 \la M/ \mathrm{M}_\odot \la 8$, because the small additional mass of hydrogen does not contribute to the output luminosity.

\begin{figure}
\includegraphics[width=0.47\textwidth]{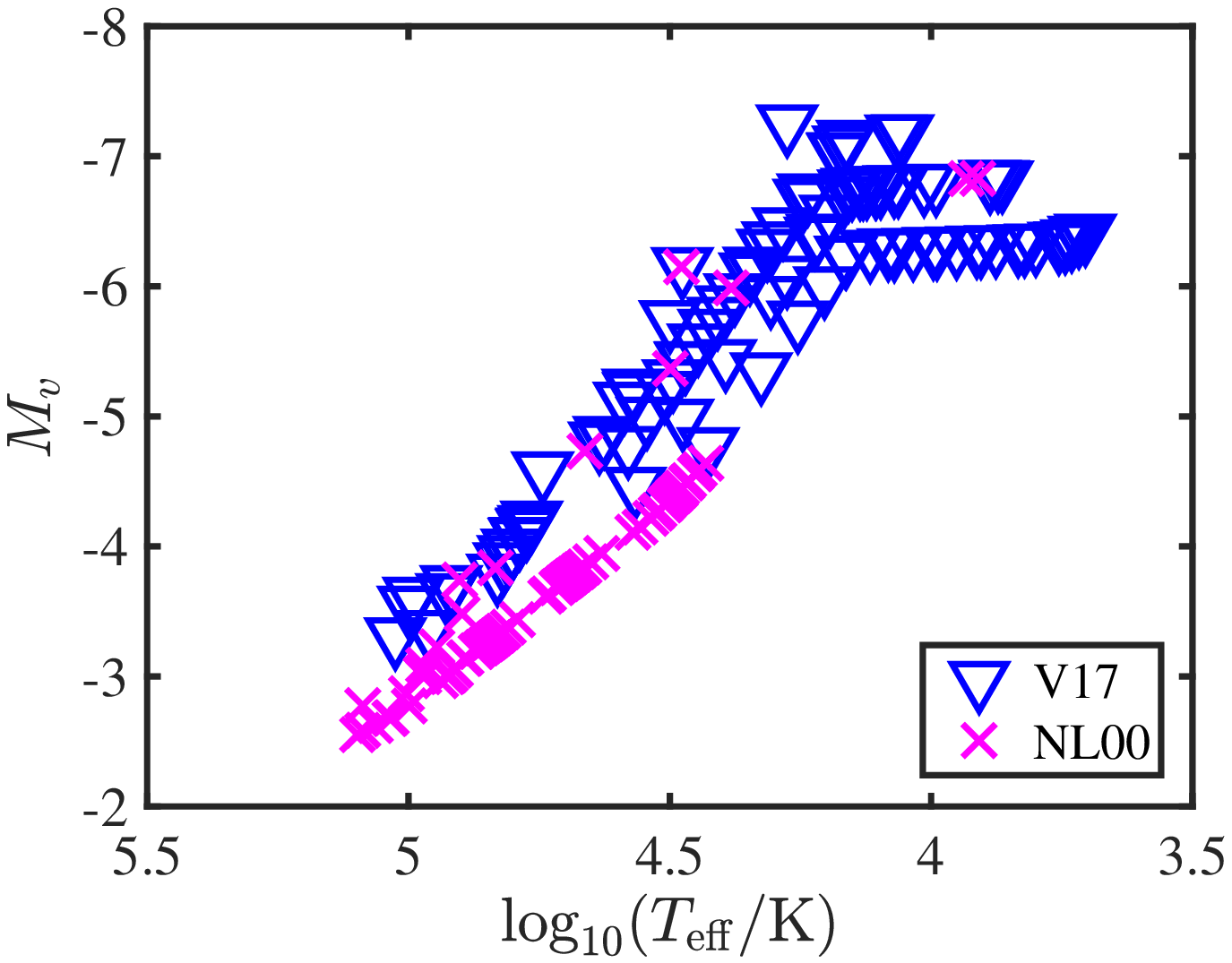}
\caption{Absolute narrow-band visual magnitude $M_v$ as function of effective surface temperature for all models with $X_\mathrm{s} < 0.4$.}
\label{fig:avmvsT}
\end{figure}
Fig. \ref{fig:avmvsT} shows the absolute narrow-band visual magnitude $M_v$ of the final models estimated, as by \cite{Yoonetal2012}, with a bolometric correction of $\mathrm{BC} = 22.053 - 5.306 \log_{10} \left( T_\mathrm{eff} / \mathrm{K} \right)$ for $T_\mathrm{eff}>14\, 330\, \mathrm{K}$ and $\mathrm{BC}=0$ for $T_\mathrm{eff} \le 14\, 330\, \mathrm{K}$. This a rudimentary approximation and is not based on detailed atmosphere models (e.g. \citealt{Eldridgeetal2017}). However, it is sufficient to indicate that less radiation is emitted in the narrow visual-band for the hotter models, because almost all the NL00 models have visual magnitudes fainter than $-5$, while the V17 models are mostly brighter. The lower mass models tend to be even brighter than the WR stars, in terms of $M_v$, prior to explosion. In some cases, though these are the minority, the secondary is brighter than the primary. These are mostly stars which overflow their Roche lobe again after a late expansion phase. There are more such cases for the V17 models which expand significantly more than the NL00 models (Fig. \ref{fig:RvsM}).

Further mass transfer by RLOF after the simulation ends is not expected to change the properties of the models as CCSN progenitors very much because the few years until core collapse do not allow for a significant change in the envelope mass, even when it is already quite small. However, some models are already affected by a late mass-transfer phase which begins long before core carbon depletion and is included in the simulation. This has the effect, for example, of limiting the photosphere size to the Roche lobe radius. To assess the importance of this effect, we evolved the same stars from core helium depletion without companions. The results of these additional runs show that individual cases then expand more and have cooler effective temperatures, most notably the systems with initial primary masses of $M_1=12$ and $14\,\mathrm{M}_\odot$ for which the radius can be an order of magnitude larger and the temperature a factor of $2$ lower, but the overall range of stellar properties, such as temperatures and radii, are unchanged. The total mass of hydrogen left in the envelope is affected by the late mass-transfer phase because mass is lost through RLOF as well as by stellar winds. The minimal hydrogen mass for the V17 models, evolved as single stars after core helium depletion, is $0.005\,\mathrm{M}_\odot$, somewhat greater than the $0.0008\,\mathrm{M}_\odot$ for the models which include the late mass-transfer phase (Fig.~\ref{fig:TvsMH}). For those with initial masses of $M_1=12$ and $14\,\mathrm{M}_\odot$, $M_\mathrm{H} > 0.018\,\mathrm{M}_\odot$. These modest quantitative differences do not substantially affect our main conclusions.

The mass-transfer rates for models which experience late RLOF are typically $\dot{M} \approx 10^{-5} \times \mathrm{M}_\odot \, \mathrm{yr}^{-1}$. This is an order-of-magnitude lower than the rates of about $ 10^{-4} \times \mathrm{M}_\odot \, \mathrm{yr}^{-1}$ given by \cite*{Taurisetal2015} because they have mass transfer from a helium star onto a less massive neutron star, while in our models the late mass transfer is always from a hydrogen-poor star on to a more massive companion, because the earlier mass-transfer episode inverted the mass ratio\footnote{Mass transfer on to a more massive companion results in a widening of the orbit and a lower mass-transfer rate compared to the case when the companion is the less massive star.}. Further details for all our models are given in Appendix \ref{app:models}.

\section{SUMMARY AND DISCUSSION}
\label{sec:summary}

We find that the retention of hydrogen, in the primary of a massive binary system, is highly sensitive to the assumed stellar wind mass-loss rate after RLOF. The two different mass-loss rates used in our study (\citealt{NL00} and \citealt{V17}) give rise to potential CCSN progenitors with very different characteristics. Almost all models which employed the NL00 winds lost all of their hydrogen, while models with the V17 mass-loss prescription did not. These results are of course metallicity-dependent because mass loss by line-driven winds depends on the chemical abundances in the photosphere, as is evident from the dependence on $Z_\mathrm{s}$ in equations (\ref{eq:MdotV17}) and (\ref{eq:MdotNL00}).

The evolutionary end points of the V17 models also tended toward lower temperatures, larger photospheric radii and to surface helium abundances covering a wide range up to about 0.9. The NL00 models almost all have a helium surface abundance of $Y_\mathrm{s} \ga 0.98$ because no hydrogen is left. Acknowledging the uncertainties in modelling SNe spectra and light curves, we can cautiously say that use of the V17 mass-loss rate instead of the NL00 shifts binary progenitor models for CCSNe over a large initial parameter space from Type~Ib to Type~IIb. For lower metallicites the mass-loss rate is expected to be smaller so there would be even more SNe of Type~IIb relative to SNe of Type~Ib, as pointed out by \cite{Yoonetal2017}.

The V17 models in our study are mostly brighter in the visual than our NL00 models (Fig. \ref{fig:avmvsT}). It would also be hard to reconcile the V17 models, which are mostly quite visually bright (low $M_v$), with the detection limits of Type~Ib SNe \citep{Eldridgeetal2013,McClellandEldridge2016}. It is likely that stripped stars have lower wind mass-loss rates than given by NL00 and, with our experiment with the V17 rate, it seems as if stable mass-transfer leads to more Type~IIb SNe than Type~Ib SNe but we do not yet have a definitive statement. A statistical analysis is needed to compare binary evolution models with the overall rates of different types of CCSNe \citep{Smithetal2011,Grauretal2017}. 

The absence of known analogues to our suggested hydrogen-poor giant Type~IIb SNe progenitors is puzzling. At high temperatures, such as most of our NL00 models, the primary stars can remain hidden by their companions because most of the luminosity is output in the far ultraviolet \citep{Gotbergetal2018}. Our V17 models with lower mass-loss rates should at some point in their evolution be significantly cooler and more visible. So we would expect to see more such stars in the Milky Way or the Local Group. One relevant example is the helium giant $\upsilon$~Sagittarii \citep{Ups1983,Ups1990,Ups2012}.

In another set of models we changed the mixing assumptions described in Section~\ref{subsec:mix} and the Schwarzschild criterion and step overshooting were used. Several more models with the NL00 prescription in this set retained a hydrogen envelope but these were still the minority. The V17 models were unaffected. This is a consequence of the dependence on the helium fraction in the NL00 prescription. This does not exist in the V17 prescription. While the qualitative results are not affected by the mixing assumptions, quantitatively the rates of Type~Ib and Type~IIb SNe can change, indicating another sensitivity to an uncertain process which affects stellar modelling.

Our study has further implications for a number of issues.

($i$)
\cite{Sravanetal2018} find it difficult to account for the rate of Type~IIb SNe at solar metallicity. They note that lower mass-loss rates would alleviate the situation. Our findings strongly support this idea, though a rigorous statistical analysis is warranted, specifically to address the impact on Type~Ib rates and to compare with observed rates.

($ii$)
Helium giant stars with final masses in the range $2 < M /  \mathrm{M}_\odot < 4$ have been suggested as possible progenitors of rapidly-fading supernovae \citep*{Kleiseretal2018}. Our stellar models share many similarities with those of \cite{Kleiseretal2018} even when the stars retain some hydrogen. These might also be relevant for rapidly-fading supernovae. Wind mass loss in helium stars is similarly important for electron-capture SNe in binary systems \citep{Taurisetal2015,MoriyaEldridge2016}.

($iii$)
The implications for the ionizing radiation provided by massive stars which have lost their envelopes by RLOF need to be assessed (\citealt{Stanwayetal2016,Gotbergetal2017,Xiaoetal2018}). Our models with the V17 mass-loss rate reach hot UV-producing regions in the Hertzsprung--Russell diagram during part of their evolution (see Fig. \ref{fig:HR1}) but do not get as hot as the models with the NL00 mass-loss rate.

We end by reiterating that the main point of this study is to illustrate the sensitivity of evolutionary models for CCSN progenitors and the need for a sound theoretical understanding of stellar winds.

\section*{Acknowledgments}
The authors thank an anonymous referee for constructive comments. AG gratefully acknowledges the support of the Blavatnik Family Foundation. JJE acknowledges travel support from the University of Auckland. CAT thanks Churchill College for his fellowship.

\bibliographystyle{mnras}

\appendix
\section{Code implementation}
\label{app:code}

Here we list some specific details of the \textsc{mesa} implementation.

The wind mass-loss rate was calculated with the \texttt{other\_wind} hook in the \texttt{MESA run\_star\_extras.f} file. An input parameter (\texttt{x\_character\_ctrl}) was used to distinguish between the NL00 and V17 schemes. Except for the hot hydrogen-deficient phase of models which employed the V17 prescription, the mass-loss rate was similar to the Dutch scheme of \textsc{mesa}.

The mass-transfer efficiency was limited according to two criteria. The first is equation (\ref{eq:Mdotth}) which gives
\begin{eqnarray}
\begin{aligned}
\beta_\mathrm{th} = \frac{1}{\tau_\mathrm{th}} \frac{M}{\dot{M}_\mathrm{tr}}.
\label{eq:betath}
\end{aligned}
\end{eqnarray}
The second criterion is related to the radius of the accretor compared to its Roche lobe radius. This gives
\begin{eqnarray}
\beta_\mathrm{L}=
\begin{cases}
    1,& R_2 \le 0.99 R_{\mathrm{L},2},\\
    f_3\left(R_2/R_{\mathrm{L},2}\right),& 0.99 R_{\mathrm{L},2}<R_2<R_{\mathrm{L},2},\\
    0,              & R_2 \ge R_{\mathrm{L},2},
\end{cases}
\label{eq:betarl}
\end{eqnarray}
with
\begin{eqnarray}
\begin{aligned}
f_3\left(x\right)=10^6 \left(2 x^3 - 5.97 x^2 +5.94 x - 1.97 \right).
\label{eq:poly3}
\end{aligned}
\end{eqnarray}
The two criteria are combined as
\begin{eqnarray}
\begin{aligned}
\beta_\mathrm{max} = \min \left( \beta_\mathrm{th} , \beta_\mathrm{L} , 0.9 \right).
\label{eq:betamax}
\end{aligned}
\end{eqnarray}
The enforcement of equation (\ref{eq:betamax}) was made by the function \texttt{extras\_binary\_check\_model} in the \texttt{MESA run\_binary\_extras.f} file. Whenever $\beta$ used for a time step deviates from that calculated by equation (\ref{eq:betamax}) with given $M$, $L$, $R$, $R_2$ and $R_{\mathrm{L},2}$ at the end of the time step by more than $\Delta \beta$ (chosen as $\Delta \beta=0.001$) \texttt{extras\_binary\_check\_model} tells the code to rerun the time step with a different $\beta$, chosen in an informed manner. This is iterated until convergence in a similar way to the implicit method of mass transfer described by \cite{Paxton2015}. We note that the mass-transfer rate itself $\dot{M}_\mathrm{tr}$ is computed explicitly from the stellar parameters at the beginning of the time step.

In addition to the mixing described in Section~\ref{subsec:mix}, the outermost part of the accretor had enhanced mixing, implemented with the \texttt{other\_D\_mix} hook. The part for enhanced mixing was chosen as the region defined by $0.99<m/M_2<1$, where $m$ is the mass coordinate within the accretor and $M_2$ is the total mass of the accretor. In this region the mixing coefficient is set to $D_\mathrm{mix,out}=10^{20} \, \mathrm{cm}^2 \, \mathrm{s}^{-1}$ but only if the star is gaining mass. This mixing enhancement is added because the accretion of material with a composition significantly different from the surface composition of the accretor causes abrupt changes in the surface opacity and radius, and related quantities. The enhanced mixing ensures a smooth evolution of the secondary during accretion of helium-rich material. Because our focus is on the properties of the primary, which are anyway rather insensitive to the details of the secondary, this modification is of minor importance.

\section{Stellar models}
\label{app:models}

The initial parameters we used give a total of 198 different combinations. Of these 93 never reached a point in their evolution at which $X_\mathrm{s}<0.4$ and are not discussed or presented (except for the $M_1=12\,\mathrm{M}_\odot$, $M_2=10\,\mathrm{M}_\odot$, $P_\mathrm{i}=367\,\mathrm{d}$ track shown in Fig. \ref{fig:HR1}). The remaining 105 combinations of initial parameters become 210 separate evolutionary tracks because different mass-loss recipes are used once $X_\mathrm{s}<0.4$. Of these 3 (listed in Table \ref{tab:convprob}) have convergence problems and 14 (listed in Table \ref{tab:cee}) head towards common envelope evolution. The properties of interest for the remaining 193 binary systems modelled are listed in Tables \ref{tab:modelsV17a}--\ref{tab:modelsNL00c}, where $f_\mathrm{L}=\left(R-R_{\mathrm{L},1}\right)/R_{\mathrm{L},1}$, where $R_{\mathrm{L},1}$ is the Roche lobe radius of the primary. The mass-loss rate given in the last column is the sum of the wind mass-loss rate and mass-transfer rate by RLOF. The other properties listed in the tables have been defined earlier. These 193 models, all of which reached central carbon depletion, are presented and discussed throughout the paper.
\begin{table}
\centering
\caption{Initial parameters for which convergence problems arise after reaching $X_\mathrm{s}<0.4$.}
\begin{threeparttable}
\begin{tabular}{cccc}
\hline
$M_1/$ & $M_2/$ & $P_\mathrm{i}/$ & mass-loss recipe\\
$\mathrm{M}_\odot$ & $\mathrm{M}_\odot$ & $\mathrm{d}$ & \\
\hline
$12$ & $5$ & $5$ & NL00 \\
$16$ & $10$ & $33$ & NL00 \\
$22$ & $8$ & $10$ & V17 \\
\hline
\hline
\end{tabular}
\footnotesize
\label{Table:ConvergenceProblems}
\end{threeparttable}
\label{tab:convprob}
\end{table}
\begin{table}
\centering
\caption{Initial parameters for which the evolution headed toward common envelope evolution after reaching $X_\mathrm{s}<0.4$, regardless of the mass-loss recipe employed.}
\begin{threeparttable}
\begin{tabular}{ccc}
\hline
$M_1/$ & $M_2/$ & $P_\mathrm{i}/$\\
$\mathrm{M}_\odot$ & $\mathrm{M}_\odot$ & $\mathrm{d}$\\
\hline
$12$ & $5$ & $1219$ \\
$14$ & $5$ & $669$ \\
$14$ & $5$ & $1219$ \\
$16$ & $6$ & $669$ \\
$16$ & $6$ & $1219$ \\
$16$ & $6$ & $2223$ \\
$19$ & $7$ & $1219$ \\
\hline
\hline
\end{tabular}
\footnotesize
\label{Table:CommonEnvelope}
\end{threeparttable}
\label{tab:cee}
\end{table}
\begin{table*}
\centering
\caption{Initial parameters and final properties for stellar evolution calculations with $0.9>q>0.8$ with the V17 prescription.}
\begin{threeparttable}
\begin{tabular}{ccc|cccccccccc}
\hline
$M_1/$ & $M_2/$ & $P_\mathrm{i}/$ & $M_\mathrm{H}/$ & $M/$ & $R/$ & $T_\mathrm{eff}/$ & $L/$ & $M_v$ & $X_\mathrm{s}$ & $Y_\mathrm{s}$ & $f_\mathrm{L}$ &$\log_{10} \left( | \dot{M} | /  \mathrm{M}_\odot \, \mathrm{yr}^{-1}\right)$\\
$\mathrm{M}_\odot$ & $\mathrm{M}_\odot$ & $\mathrm{d}$ & $\mathrm{M}_\odot$ & $\mathrm{M}_\odot$ & $\mathrm{R}_\odot$ & $\mathrm{K}$ & $\mathrm{L}_\odot$ & & & & & \\
\hline
$12$ & $10$ & $5$ & $0.0075$ & $3.09$ & $35.49$ & $14538$ & $50539$ & $-6.2$ & $0.22$ & $0.76$ & $0.033$ & $-5.29$\\
$12$ & $10$ & $10$ & $0.0101$ & $3.14$ & $54.81$ & $11760$ & $51616$ & $-6.3$ & $0.23$ & $0.75$ & $0.023$ & $-5.3$\\
$12$ & $10$ & $18$ & $0.012$ & $3.16$ & $80.3$ & $9744$ & $52219$ & $-6.3$ & $0.24$ & $0.74$ & $0.019$ & $-5.28$\\
$12$ & $10$ & $33$ & $0.0146$ & $3.19$ & $118.9$ & $8029$ & $52786$ & $-6.3$ & $0.25$ & $0.73$ & $0.008$ & $-5.29$\\
$12$ & $10$ & $60$ & $0.0183$ & $3.22$ & $175.53$ & $6627$ & $53373$ & $-6.3$ & $0.26$ & $0.72$ & $-0.013$ & $-5.29$\\
$12$ & $10$ & $110$ & $0.0272$ & $3.29$ & $253.89$ & $5539$ & $54531$ & $-6.4$ & $0.27$ & $0.71$ & $-0.038$ & $-5.4$\\
$12$ & $10$ & $201$ & $0.0359$ & $3.37$ & $291.96$ & $5216$ & $56679$ & $-6.4$ & $0.31$ & $0.67$ & $-0.224$ & $-5.98$\\
$14$ & $12$ & $5$ & $0.0248$ & $4.05$ & $33.45$ & $16671$ & $77654$ & $-6.4$ & $0.24$ & $0.74$ & $0.064$ & $-5.05$\\
$14$ & $12$ & $10$ & $0.029$ & $4.11$ & $43.54$ & $14686$ & $79225$ & $-6.7$ & $0.25$ & $0.73$ & $0.051$ & $-6.64$\\
$14$ & $12$ & $18$ & $0.0318$ & $4.13$ & $50.43$ & $13676$ & $79932$ & $-6.8$ & $0.26$ & $0.72$ & $0.007$ & $-6.64$\\
$14$ & $12$ & $33$ & $0.0331$ & $4.16$ & $58.31$ & $12741$ & $80497$ & $-6.8$ & $0.27$ & $0.72$ & $-0.209$ & $-6.64$\\
$14$ & $12$ & $60$ & $0.0345$ & $4.18$ & $68.6$ & $11771$ & $81164$ & $-6.8$ & $0.27$ & $0.71$ & $-0.37$ & $-6.63$\\
$14$ & $12$ & $110$ & $0.0362$ & $4.21$ & $89.65$ & $10317$ & $81798$ & $-6.8$ & $0.28$ & $0.7$ & $-0.487$ & $-6.29$\\
$14$ & $12$ & $201$ & $0.0413$ & $4.28$ & $162.01$ & $7712$ & $83412$ & $-6.8$ & $0.3$ & $0.68$ & $-0.474$ & $-5.97$\\
$14$ & $12$ & $367$ & $0.0421$ & $4.3$ & $174.53$ & $7446$ & $84135$ & $-6.8$ & $0.31$ & $0.68$ & $-0.603$ & $-5.94$\\
$16$ & $14$ & $5$ & $0.0312$ & $4.96$ & $12.62$ & $29448$ & $107581$ & $-5.4$ & $0.25$ & $0.73$ & $-0.517$ & $-6.46$\\
$16$ & $14$ & $10$ & $0.0352$ & $5.03$ & $16.83$ & $25605$ & $109411$ & $-5.8$ & $0.27$ & $0.71$ & $-0.572$ & $-6.45$\\
$16$ & $14$ & $110$ & $0.044$ & $5.15$ & $36.52$ & $17516$ & $112776$ & $-6.7$ & $0.29$ & $0.69$ & $-0.802$ & $-6.44$\\
$16$ & $14$ & $201$ & $0.0472$ & $5.2$ & $51.9$ & $14729$ & $113919$ & $-7.1$ & $0.3$ & $0.68$ & $-0.82$ & $-6.43$\\
$16$ & $14$ & $367$ & $0.0467$ & $5.19$ & $49.38$ & $15098$ & $113841$ & $-7$ & $0.3$ & $0.68$ & $-0.882$ & $-6.43$\\
$16$ & $14$ & $669$ & $0.0516$ & $5.27$ & $88.74$ & $11323$ & $116305$ & $-7.2$ & $0.31$ & $0.67$ & $-0.848$ & $-6.42$\\
$19$ & $16$ & $5$ & $0.0143$ & $5.86$ & $3.04$ & $64430$ & $143389$ & $-3.9$ & $0.14$ & $0.84$ & $-0.89$ & $-6.29$\\
$19$ & $16$ & $10$ & $0.0256$ & $5.73$ & $6.63$ & $43133$ & $136858$ & $-4.8$ & $0.21$ & $0.77$ & $-0.858$ & $-6.32$\\
$19$ & $16$ & $18$ & $0.026$ & $5.95$ & $9.19$ & $37194$ & $145246$ & $-5.2$ & $0.36$ & $0.63$ & $-0.866$ & $-6.28$\\
$19$ & $16$ & $33$ & $0.0181$ & $5.95$ & $3.39$ & $61363$ & $146135$ & $-4.1$ & $0.15$ & $0.83$ & $-0.965$ & $-6.28$\\
$22$ & $19$ & $5$ & $0.013$ & $7.23$ & $2.16$ & $82848$ & $197679$ & $-3.7$ & $0.13$ & $0.85$ & $-0.925$ & $-6.11$\\
$22$ & $19$ & $10$ & $0.0304$ & $7.07$ & $4.72$ & $55393$ & $188335$ & $-4.6$ & $0.21$ & $0.77$ & $-0.902$ & $-6.13$\\
$25$ & $22$ & $5$ & $0.0124$ & $8.69$ & $1.78$ & $97740$ & $258873$ & $-3.6$ & $0.13$ & $0.85$ & $-0.94$ & $-5.95$\\
\hline
\hline
\end{tabular}
\footnotesize
\label{Table:Models1}
\end{threeparttable}
\label{tab:modelsV17a}
\end{table*}
\begin{table*}
\centering
\caption{Initial parameters and final properties for stellar evolution calculations with $0.9>q>0.8$ with the NL00 prescription.}
\begin{threeparttable}
\begin{tabular}{ccc|cccccccccc}
\hline
$M_1/$ & $M_2/$ & $P_\mathrm{i}/$ & $M_\mathrm{H}/$ & $M/$ & $R/$ & $T_\mathrm{eff}/$ & $L/$ & $M_v$ & $X_\mathrm{s}$ & $Y_\mathrm{s}$ & $f_\mathrm{L}$ &$\log_{10} \left( | \dot{M} | /  \mathrm{M}_\odot \, \mathrm{yr}^{-1}\right)$\\
$\mathrm{M}_\odot$ & $\mathrm{M}_\odot$ & $\mathrm{d}$ & $\mathrm{M}_\odot$ & $\mathrm{M}_\odot$ & $\mathrm{R}_\odot$ & $\mathrm{K}$ & $\mathrm{L}_\odot$ & & & & & \\
\hline
$12$ & $10$ & $5$ & $0$ & $2.79$ & $8.76$ & $28256$ & $43960$ & $-4.6$ & $0$ & $0.98$ & $-0.461$ & $-5.89$\\
$12$ & $10$ & $10$ & $0$ & $2.82$ & $8.19$ & $29378$ & $44898$ & $-4.5$ & $0$ & $0.98$ & $-0.721$ & $-5.88$\\
$12$ & $10$ & $18$ & $0$ & $2.85$ & $7.67$ & $30466$ & $45497$ & $-4.4$ & $0$ & $0.98$ & $-0.836$ & $-5.87$\\
$12$ & $10$ & $33$ & $0$ & $2.87$ & $7.38$ & $31150$ & $46053$ & $-4.4$ & $0$ & $0.98$ & $-0.9$ & $-5.87$\\
$12$ & $10$ & $60$ & $0$ & $2.89$ & $7.02$ & $32026$ & $46644$ & $-4.3$ & $0$ & $0.98$ & $-0.94$ & $-5.86$\\
$12$ & $10$ & $110$ & $0$ & $2.94$ & $6.48$ & $33569$ & $47964$ & $-4.3$ & $0$ & $0.98$ & $-0.966$ & $-5.84$\\
$12$ & $10$ & $201$ & $0$ & $3.03$ & $5.77$ & $36062$ & $50593$ & $-4.1$ & $0$ & $0.98$ & $-0.981$ & $-5.81$\\
$14$ & $12$ & $5$ & $0$ & $3.48$ & $3.81$ & $47440$ & $66065$ & $-3.8$ & $0$ & $0.98$ & $-0.865$ & $-5.66$\\
$14$ & $12$ & $10$ & $0$ & $3.53$ & $3.66$ & $48683$ & $67612$ & $-3.8$ & $0$ & $0.98$ & $-0.915$ & $-5.65$\\
$14$ & $12$ & $18$ & $0$ & $3.55$ & $3.58$ & $49344$ & $68431$ & $-3.8$ & $0$ & $0.98$ & $-0.944$ & $-5.64$\\
$14$ & $12$ & $33$ & $0$ & $3.58$ & $3.52$ & $49954$ & $69268$ & $-3.7$ & $0$ & $0.98$ & $-0.964$ & $-5.64$\\
$14$ & $12$ & $60$ & $0$ & $3.61$ & $3.44$ & $50714$ & $70147$ & $-3.7$ & $0$ & $0.98$ & $-0.977$ & $-5.63$\\
$14$ & $12$ & $110$ & $0$ & $3.64$ & $3.34$ & $51682$ & $71357$ & $-3.7$ & $0$ & $0.98$ & $-0.985$ & $-5.62$\\
$14$ & $12$ & $201$ & $0$ & $3.73$ & $3.12$ & $53966$ & $74097$ & $-3.6$ & $0$ & $0.98$ & $-0.991$ & $-5.6$\\
$14$ & $12$ & $367$ & $0.0375$ & $4.29$ & $132.02$ & $8562$ & $84167$ & $-6.8$ & $0.3$ & $0.68$ & $-0.688$ & $-6.04$\\
$16$ & $14$ & $5$ & $0$ & $4.15$ & $2.22$ & $66826$ & $88172$ & $-3.3$ & $0$ & $0.98$ & $-0.921$ & $-5.5$\\
$16$ & $14$ & $10$ & $0$ & $4.2$ & $2.14$ & $68460$ & $89973$ & $-3.3$ & $0$ & $0.98$ & $-0.952$ & $-5.49$\\
$16$ & $14$ & $110$ & $0$ & $4.35$ & $1.96$ & $72360$ & $94599$ & $-3.2$ & $0$ & $0.98$ & $-0.991$ & $-5.46$\\
$16$ & $14$ & $201$ & $0$ & $4.41$ & $1.9$ & $73946$ & $96637$ & $-3.2$ & $0$ & $0.98$ & $-0.994$ & $-5.45$\\
$16$ & $14$ & $367$ & $0$ & $4.57$ & $1.77$ & $77640$ & $102082$ & $-3.1$ & $0$ & $0.98$ & $-0.996$ & $-5.42$\\
$16$ & $14$ & $669$ & $0.0308$ & $5.2$ & $19.64$ & $24117$ & $117223$ & $-6$ & $0.3$ & $0.68$ & $-0.969$ & $-5.61$\\
$19$ & $16$ & $5$ & $0$ & $4.68$ & $1.53$ & $84205$ & $106060$ & $-3$ & $0$ & $0.98$ & $-0.947$ & $-5.4$\\
$19$ & $16$ & $10$ & $0$ & $4.6$ & $1.62$ & $81333$ & $103510$ & $-3.1$ & $0$ & $0.98$ & $-0.966$ & $-5.41$\\
$19$ & $16$ & $18$ & $0.0249$ & $5.94$ & $6.01$ & $45968$ & $145061$ & $-4.7$ & $0.23$ & $0.75$ & $-0.91$ & $-5.42$\\
$19$ & $16$ & $33$ & $0.0039$ & $5.83$ & $2.04$ & $78963$ & $144673$ & $-3.5$ & $0.11$ & $0.87$ & $-0.98$ & $-5.31$\\
$22$ & $19$ & $5$ & $0$ & $5.53$ & $1.04$ & $108497$ & $135133$ & $-2.7$ & $0$ & $0.98$ & $-0.963$ & $-5.26$\\
$22$ & $19$ & $10$ & $0.0102$ & $6.95$ & $2.27$ & $79799$ & $187614$ & $-3.7$ & $0.16$ & $0.82$ & $-0.953$ & $-5.21$\\
$25$ & $22$ & $5$ & $0$ & $6.38$ & $0.86$ & $125125$ & $163105$ & $-2.6$ & $0$ & $0.98$ & $-0.97$ & $-5.16$\\
\hline
\hline
\end{tabular}
\footnotesize
\label{Table:Models2}
\end{threeparttable}
\label{tab:modelsNL00a}
\end{table*}
\begin{table*}
\centering
\caption{Initial parameters and final properties for stellar evolution calculations with $0.7>q>0.6$ with the V17 prescription.}
\begin{threeparttable}
\begin{tabular}{ccc|cccccccccc}
\hline
$M_1/$ & $M_2/$ & $P_\mathrm{i}/$ & $M_\mathrm{H}/$ & $M/$ & $R/$ & $T_\mathrm{eff}/$ & $L/$ & $M_v$ & $X_\mathrm{s}$ & $Y_\mathrm{s}$ & $f_\mathrm{L}$ &$\log_{10} \left( | \dot{M} | /  \mathrm{M}_\odot \, \mathrm{yr}^{-1}\right)$\\
$\mathrm{M}_\odot$ & $\mathrm{M}_\odot$ & $\mathrm{d}$ & $\mathrm{M}_\odot$ & $\mathrm{M}_\odot$ & $\mathrm{R}_\odot$ & $\mathrm{K}$ & $\mathrm{L}_\odot$ & & & & & \\
\hline
$12$ & $8$ & $5$ & $0.0063$ & $3.07$ & $29.17$ & $15996$ & $50046$ & $-6$ & $0.21$ & $0.77$ & $0.042$ & $-5.37$\\
$12$ & $8$ & $10$ & $0.0089$ & $3.12$ & $44.17$ & $13080$ & $51312$ & $-6.3$ & $0.23$ & $0.75$ & $0.025$ & $-5.29$\\
$12$ & $8$ & $18$ & $0.011$ & $3.15$ & $64.94$ & $10824$ & $51995$ & $-6.3$ & $0.24$ & $0.74$ & $0.022$ & $-5.28$\\
$12$ & $8$ & $33$ & $0.0129$ & $3.17$ & $95.67$ & $8940$ & $52519$ & $-6.3$ & $0.24$ & $0.74$ & $0.017$ & $-5.29$\\
$12$ & $8$ & $60$ & $0.0156$ & $3.2$ & $135.91$ & $7518$ & $53008$ & $-6.3$ & $0.25$ & $0.73$ & $0.002$ & $-5.29$\\
$12$ & $8$ & $110$ & $0.0198$ & $3.23$ & $194.22$ & $6306$ & $53604$ & $-6.3$ & $0.26$ & $0.72$ & $-0.018$ & $-5.28$\\
$12$ & $8$ & $201$ & $0.0312$ & $3.32$ & $270.61$ & $5383$ & $55240$ & $-6.4$ & $0.28$ & $0.7$ & $-0.095$ & $-6.02$\\
$12$ & $8$ & $367$ & $0.0412$ & $3.4$ & $311.49$ & $5067$ & $57475$ & $-6.4$ & $0.35$ & $0.63$ & $-0.275$ & $-5.95$\\
$14$ & $9$ & $5$ & $0.0222$ & $4.02$ & $25.92$ & $18863$ & $76431$ & $-6.1$ & $0.24$ & $0.75$ & $0.056$ & $-4.74$\\
$14$ & $9$ & $10$ & $0.0265$ & $4.09$ & $38.47$ & $15600$ & $78757$ & $-6.6$ & $0.25$ & $0.73$ & $0.066$ & $-5.71$\\
$14$ & $9$ & $18$ & $0.0301$ & $4.12$ & $46.9$ & $14165$ & $79557$ & $-6.8$ & $0.26$ & $0.72$ & $0.035$ & $-6.64$\\
$14$ & $9$ & $33$ & $0.0323$ & $4.14$ & $53.61$ & $13274$ & $80161$ & $-6.8$ & $0.26$ & $0.72$ & $-0.052$ & $-6.64$\\
$14$ & $9$ & $60$ & $0.0335$ & $4.16$ & $60.77$ & $12487$ & $80667$ & $-6.8$ & $0.27$ & $0.71$ & $-0.235$ & $-6.63$\\
$14$ & $9$ & $110$ & $0.0347$ & $4.19$ & $71.2$ & $11555$ & $81191$ & $-6.8$ & $0.27$ & $0.71$ & $-0.393$ & $-6.63$\\
$14$ & $9$ & $201$ & $0.0368$ & $4.22$ & $99.71$ & $9786$ & $81919$ & $-6.8$ & $0.28$ & $0.7$ & $-0.503$ & $-6.16$\\
$14$ & $9$ & $367$ & $0.0429$ & $4.3$ & $180.91$ & $7308$ & $83853$ & $-6.8$ & $0.31$ & $0.68$ & $-0.469$ & $-5.92$\\
$14$ & $9$ & $669$ & $0.0405$ & $4.3$ & $162.31$ & $7725$ & $84276$ & $-6.8$ & $0.31$ & $0.67$ & $-0.653$ & $-5.96$\\
$16$ & $10$ & $5$ & $0.0289$ & $4.93$ & $11.22$ & $31160$ & $106634$ & $-5.3$ & $0.24$ & $0.74$ & $-0.445$ & $-6.47$\\
$16$ & $10$ & $10$ & $0.033$ & $5$ & $14.13$ & $27901$ & $108698$ & $-5.6$ & $0.26$ & $0.72$ & $-0.52$ & $-6.46$\\
$16$ & $10$ & $18$ & $0.0364$ & $5.05$ & $18.12$ & $24708$ & $109964$ & $-5.9$ & $0.27$ & $0.71$ & $-0.571$ & $-6.45$\\
$16$ & $10$ & $33$ & $0.039$ & $5.08$ & $22.5$ & $22223$ & $110913$ & $-6.1$ & $0.28$ & $0.7$ & $-0.636$ & $-6.45$\\
$16$ & $10$ & $60$ & $0.0407$ & $5.11$ & $26.3$ & $20584$ & $111586$ & $-6.3$ & $0.28$ & $0.7$ & $-0.702$ & $-6.44$\\
$16$ & $10$ & $110$ & $0.0422$ & $5.13$ & $30.3$ & $19197$ & $112028$ & $-6.5$ & $0.29$ & $0.69$ & $-0.762$ & $-6.44$\\
$16$ & $10$ & $201$ & $0.0443$ & $5.16$ & $37.85$ & $17211$ & $112946$ & $-6.7$ & $0.29$ & $0.69$ & $-0.808$ & $-6.44$\\
$16$ & $10$ & $367$ & $0.0476$ & $5.2$ & $53.86$ & $14464$ & $114094$ & $-7.1$ & $0.3$ & $0.68$ & $-0.822$ & $-6.43$\\
$16$ & $10$ & $669$ & $0.0467$ & $5.19$ & $49.63$ & $15064$ & $113933$ & $-7$ & $0.3$ & $0.68$ & $-0.884$ & $-6.43$\\
$16$ & $10$ & $1219$ & $0.0512$ & $5.26$ & $83.3$ & $11682$ & $116073$ & $-7.2$ & $0.31$ & $0.67$ & $-0.853$ & $-6.42$\\
$19$ & $12$ & $5$ & $0.0132$ & $5.81$ & $2.92$ & $65627$ & $141914$ & $-3.9$ & $0.14$ & $0.85$ & $-0.867$ & $-6.3$\\
$19$ & $12$ & $10$ & $0.0262$ & $5.71$ & $6.9$ & $42240$ & $136163$ & $-4.9$ & $0.22$ & $0.76$ & $-0.805$ & $-6.32$\\
$19$ & $12$ & $18$ & $0.0262$ & $5.91$ & $8.47$ & $38644$ & $143652$ & $-5.1$ & $0.27$ & $0.72$ & $-0.837$ & $-6.29$\\
$19$ & $12$ & $33$ & $0.0182$ & $5.95$ & $3.41$ & $61164$ & $146172$ & $-4.1$ & $0.15$ & $0.83$ & $-0.956$ & $-6.28$\\
$22$ & $14$ & $5$ & $0.0116$ & $7.16$ & $2.06$ & $84502$ & $194813$ & $-3.7$ & $0.12$ & $0.86$ & $-0.906$ & $-6.11$\\
$25$ & $16$ & $5$ & $0.0107$ & $8.59$ & $1.68$ & $100070$ & $255079$ & $-3.6$ & $0.12$ & $0.86$ & $-0.928$ & $-5.95$\\
$25$ & $16$ & $10$ & $0.0454$ & $7.99$ & $44.33$ & $18860$ & $223347$ & $-7.3$ & $0.36$ & $0.62$ & $0.064$ & $-4.26$\\
\hline
\hline
\end{tabular}
\footnotesize
\label{Table:Models3}
\end{threeparttable}
\label{tab:modelsV17b}
\end{table*}
\begin{table*}
\centering
\caption{Initial parameters and final properties for stellar evolution calculations with $0.7>q>0.6$ with the NL00 prescription.}
\begin{threeparttable}
\begin{tabular}{ccc|cccccccccc}
\hline
$M_1/$ & $M_2/$ & $P_\mathrm{i}/$ & $M_\mathrm{H}/$ & $M/$ & $R/$ & $T_\mathrm{eff}/$ & $L/$ & $M_v$ & $X_\mathrm{s}$ & $Y_\mathrm{s}$ & $f_\mathrm{L}$ &$\log_{10} \left( | \dot{M} | /  \mathrm{M}_\odot \, \mathrm{yr}^{-1}\right)$\\
$\mathrm{M}_\odot$ & $\mathrm{M}_\odot$ & $\mathrm{d}$ & $\mathrm{M}_\odot$ & $\mathrm{M}_\odot$ & $\mathrm{R}_\odot$ & $\mathrm{K}$ & $\mathrm{L}_\odot$ & & & & & \\
\hline
$12$ & $8$ & $5$ & $0$ & $2.78$ & $9.04$ & $27760$ & $43600$ & $-4.6$ & $0$ & $0.98$ & $-0.271$ & $-5.9$\\
$12$ & $8$ & $10$ & $0$ & $2.82$ & $8.19$ & $29338$ & $44645$ & $-4.5$ & $0$ & $0.98$ & $-0.633$ & $-5.88$\\
$12$ & $8$ & $18$ & $0$ & $2.84$ & $7.87$ & $30037$ & $45269$ & $-4.4$ & $0$ & $0.98$ & $-0.79$ & $-5.87$\\
$12$ & $8$ & $33$ & $0$ & $2.86$ & $7.46$ & $30937$ & $45794$ & $-4.4$ & $0$ & $0.98$ & $-0.871$ & $-5.87$\\
$12$ & $8$ & $60$ & $0$ & $2.87$ & $7.26$ & $31434$ & $46276$ & $-4.4$ & $0$ & $0.98$ & $-0.918$ & $-5.86$\\
$12$ & $8$ & $110$ & $0$ & $2.9$ & $6.92$ & $32310$ & $46885$ & $-4.3$ & $0$ & $0.98$ & $-0.949$ & $-5.86$\\
$12$ & $8$ & $201$ & $0$ & $2.96$ & $6.22$ & $34406$ & $48739$ & $-4.2$ & $0$ & $0.98$ & $-0.973$ & $-5.83$\\
$12$ & $8$ & $367$ & $0$ & $3.09$ & $5.5$ & $37212$ & $52128$ & $-4.1$ & $0$ & $0.98$ & $-0.985$ & $-5.8$\\
$14$ & $9$ & $5$ & $0$ & $3.46$ & $3.88$ & $46912$ & $65369$ & $-3.8$ & $0$ & $0.98$ & $-0.811$ & $-5.67$\\
$14$ & $9$ & $10$ & $0$ & $3.51$ & $3.71$ & $48225$ & $67032$ & $-3.8$ & $0$ & $0.98$ & $-0.889$ & $-5.66$\\
$14$ & $9$ & $18$ & $0$ & $3.54$ & $3.62$ & $49005$ & $68019$ & $-3.8$ & $0$ & $0.98$ & $-0.927$ & $-5.65$\\
$14$ & $9$ & $33$ & $0$ & $3.56$ & $3.55$ & $49627$ & $68769$ & $-3.7$ & $0$ & $0.98$ & $-0.95$ & $-5.64$\\
$14$ & $9$ & $60$ & $0$ & $3.59$ & $3.49$ & $50196$ & $69465$ & $-3.7$ & $0$ & $0.98$ & $-0.967$ & $-5.64$\\
$14$ & $9$ & $110$ & $0$ & $3.61$ & $3.42$ & $50873$ & $70328$ & $-3.7$ & $0$ & $0.98$ & $-0.979$ & $-5.63$\\
$14$ & $9$ & $201$ & $0$ & $3.65$ & $3.3$ & $52032$ & $71724$ & $-3.7$ & $0$ & $0.98$ & $-0.987$ & $-5.62$\\
$14$ & $9$ & $367$ & $0$ & $3.75$ & $3.06$ & $54620$ & $74853$ & $-3.6$ & $0$ & $0.98$ & $-0.992$ & $-5.59$\\
$14$ & $9$ & $669$ & $0.0388$ & $4.29$ & $146.12$ & $8142$ & $84287$ & $-6.8$ & $0.3$ & $0.68$ & $-0.686$ & $-6$\\
$16$ & $10$ & $5$ & $0$ & $4.12$ & $2.26$ & $66065$ & $87253$ & $-3.3$ & $0$ & $0.98$ & $-0.891$ & $-5.51$\\
$16$ & $10$ & $10$ & $0$ & $4.18$ & $2.17$ & $67754$ & $89157$ & $-3.3$ & $0$ & $0.98$ & $-0.934$ & $-5.5$\\
$16$ & $10$ & $18$ & $0$ & $4.22$ & $2.11$ & $68918$ & $90444$ & $-3.3$ & $0$ & $0.98$ & $-0.957$ & $-5.49$\\
$16$ & $10$ & $60$ & $0$ & $4.28$ & $2.04$ & $70528$ & $92343$ & $-3.3$ & $0$ & $0.98$ & $-0.981$ & $-5.48$\\
$16$ & $10$ & $110$ & $0$ & $4.31$ & $1.99$ & $71583$ & $93566$ & $-3.2$ & $0$ & $0.98$ & $-0.988$ & $-5.47$\\
$16$ & $10$ & $201$ & $0$ & $4.35$ & $1.95$ & $72509$ & $94751$ & $-3.2$ & $0$ & $0.98$ & $-0.992$ & $-5.46$\\
$16$ & $10$ & $367$ & $0$ & $4.42$ & $1.89$ & $74204$ & $96958$ & $-3.2$ & $0$ & $0.98$ & $-0.995$ & $-5.45$\\
$16$ & $10$ & $669$ & $0$ & $4.59$ & $1.75$ & $78080$ & $102815$ & $-3.1$ & $0$ & $0.98$ & $-0.996$ & $-5.42$\\
$16$ & $10$ & $1219$ & $0.021$ & $5.15$ & $11.47$ & $31544$ & $117001$ & $-5.4$ & $0.29$ & $0.69$ & $-0.983$ & $-5.6$\\
$19$ & $12$ & $5$ & $0$ & $4.65$ & $1.55$ & $83569$ & $105284$ & $-3$ & $0$ & $0.98$ & $-0.93$ & $-5.4$\\
$19$ & $12$ & $10$ & $0$ & $4.61$ & $1.62$ & $81506$ & $103797$ & $-3$ & $0$ & $0.98$ & $-0.956$ & $-5.41$\\
$19$ & $12$ & $18$ & $0$ & $5.41$ & $1.39$ & $93521$ & $132565$ & $-3$ & $0$ & $0.98$ & $-0.974$ & $-5.27$\\
$19$ & $12$ & $33$ & $0.0009$ & $5.78$ & $1.63$ & $88240$ & $144636$ & $-3.2$ & $0.05$ & $0.93$ & $-0.979$ & $-5.26$\\
$22$ & $14$ & $5$ & $0$ & $5.49$ & $1.05$ & $107880$ & $134159$ & $-2.7$ & $0$ & $0.98$ & $-0.95$ & $-5.27$\\
$25$ & $16$ & $5$ & $0$ & $6.32$ & $0.89$ & $122343$ & $160372$ & $-2.6$ & $0$ & $0.98$ & $-0.959$ & $-5.17$\\
$25$ & $16$ & $10$ & $0$ & $6.89$ & $0.97$ & $122289$ & $190807$ & $-2.8$ & $0$ & $0.98$ & $-0.975$ & $-5.07$\\
\hline
\hline
\end{tabular}
\footnotesize
\label{Table:Models4}
\end{threeparttable}
\label{tab:modelsNL00b}
\end{table*}
\begin{table*}
\centering
\caption{Initial parameters and final properties for stellar evolution calculations with $0.45>q>0.35$ with the V17 prescription.}
\begin{threeparttable}
\begin{tabular}{ccc|cccccccccc}
\hline
$M_1/$ & $M_2/$ & $P_\mathrm{i}/$ & $M_\mathrm{H}/$ & $M/$ & $R/$ & $T_\mathrm{eff}/$ & $L/$ & $M_v$ & $X_\mathrm{s}$ & $Y_\mathrm{s}$ & $f_\mathrm{L}$ &$\log_{10} \left( | \dot{M} | /  \mathrm{M}_\odot \, \mathrm{yr}^{-1}\right)$\\
$\mathrm{M}_\odot$ & $\mathrm{M}_\odot$ & $\mathrm{d}$ & $\mathrm{M}_\odot$ & $\mathrm{M}_\odot$ & $\mathrm{R}_\odot$ & $\mathrm{K}$ & $\mathrm{L}_\odot$ & & & & & \\
\hline
$12$ & $5$ & $5$ & $0.0009$ & $2.96$ & $10.2$ & $26706$ & $47524$ & $-4.8$ & $0.13$ & $0.85$ & $0.043$ & $-5.59$\\
$12$ & $5$ & $10$ & $0.0034$ & $3.03$ & $16.56$ & $21141$ & $49217$ & $-5.3$ & $0.19$ & $0.8$ & $0.033$ & $-5.61$\\
$12$ & $5$ & $18$ & $0.0055$ & $3.08$ & $23.14$ & $17989$ & $50371$ & $-5.7$ & $0.21$ & $0.77$ & $0.041$ & $-5.29$\\
$12$ & $5$ & $33$ & $0.0077$ & $3.11$ & $35.44$ & $14596$ & $51231$ & $-6.2$ & $0.22$ & $0.76$ & $0.035$ & $-5.31$\\
$12$ & $5$ & $60$ & $0.01$ & $3.14$ & $52.09$ & $12079$ & $51891$ & $-6.3$ & $0.23$ & $0.75$ & $0.026$ & $-5.29$\\
$12$ & $5$ & $110$ & $0.0119$ & $3.17$ & $77.85$ & $9903$ & $52372$ & $-6.3$ & $0.24$ & $0.74$ & $0.023$ & $-5.34$\\
$12$ & $5$ & $201$ & $0.0142$ & $3.19$ & $114.62$ & $8180$ & $52840$ & $-6.3$ & $0.25$ & $0.73$ & $0.007$ & $-5.28$\\
$12$ & $5$ & $367$ & $0.0177$ & $3.22$ & $165.81$ & $6816$ & $53327$ & $-6.3$ & $0.26$ & $0.72$ & $-0.009$ & $-5.32$\\
$12$ & $5$ & $669$ & $0.0247$ & $3.27$ & $239.51$ & $5697$ & $54292$ & $-6.4$ & $0.27$ & $0.71$ & $-0.033$ & $-5.3$\\
$14$ & $5$ & $5$ & $0.0038$ & $3.75$ & $6.55$ & $36851$ & $71080$ & $-4.5$ & $0.16$ & $0.82$ & $0.06$ & $-4.8$\\
$14$ & $5$ & $10$ & $0.0108$ & $3.88$ & $10.16$ & $29861$ & $73693$ & $-5$ & $0.2$ & $0.78$ & $0.046$ & $-4.62$\\
$14$ & $5$ & $18$ & $0.016$ & $3.96$ & $14.97$ & $24727$ & $75280$ & $-5.4$ & $0.22$ & $0.76$ & $0.054$ & $-4.6$\\
$14$ & $5$ & $33$ & $0.0208$ & $4.03$ & $22.57$ & $20267$ & $77236$ & $-5.9$ & $0.24$ & $0.74$ & $0.052$ & $-4.5$\\
$14$ & $5$ & $60$ & $0.0247$ & $4.08$ & $34.08$ & $16526$ & $77834$ & $-6.4$ & $0.25$ & $0.73$ & $0.073$ & $-4.68$\\
$14$ & $5$ & $110$ & $0.0284$ & $4.11$ & $45.41$ & $14389$ & $79444$ & $-6.8$ & $0.26$ & $0.73$ & $0.057$ & $-6.59$\\
$14$ & $5$ & $201$ & $0.0318$ & $4.14$ & $51.86$ & $13490$ & $80032$ & $-6.8$ & $0.26$ & $0.72$ & $0.045$ & $-6.64$\\
$14$ & $5$ & $367$ & $0.0331$ & $4.16$ & $58$ & $12776$ & $80538$ & $-6.8$ & $0.27$ & $0.72$ & $-0.158$ & $-6.64$\\
$16$ & $6$ & $5$ & $0.0199$ & $4.75$ & $7.36$ & $37973$ & $101080$ & $-4.8$ & $0.19$ & $0.79$ & $0.096$ & $-6$\\
$16$ & $6$ & $10$ & $0.0244$ & $4.86$ & $8.82$ & $34967$ & $104471$ & $-5$ & $0.21$ & $0.77$ & $-0.157$ & $-6.48$\\
$16$ & $6$ & $18$ & $0.0275$ & $4.92$ & $10.39$ & $32367$ & $106411$ & $-5.2$ & $0.24$ & $0.74$ & $-0.327$ & $-6.47$\\
$16$ & $6$ & $33$ & $0.0312$ & $4.97$ & $12.56$ & $29539$ & $107938$ & $-5.4$ & $0.25$ & $0.73$ & $-0.422$ & $-6.46$\\
$16$ & $6$ & $60$ & $0.0344$ & $5.02$ & $15.55$ & $26623$ & $109183$ & $-5.7$ & $0.26$ & $0.72$ & $-0.495$ & $-6.46$\\
$16$ & $6$ & $110$ & $0.0374$ & $5.06$ & $19.64$ & $23752$ & $110251$ & $-6$ & $0.27$ & $0.71$ & $-0.561$ & $-6.45$\\
$16$ & $6$ & $201$ & $0.0395$ & $5.09$ & $23.36$ & $21818$ & $111105$ & $-6.2$ & $0.28$ & $0.7$ & $-0.638$ & $-6.45$\\
$16$ & $6$ & $367$ & $0.0411$ & $5.11$ & $27.61$ & $20096$ & $111701$ & $-6.4$ & $0.28$ & $0.7$ & $-0.708$ & $-6.44$\\
$19$ & $7$ & $5$ & $0.0114$ & $5.74$ & $2.73$ & $67581$ & $139282$ & $-3.8$ & $0.12$ & $0.86$ & $-0.671$ & $-6.31$\\
$19$ & $7$ & $10$ & $0.0286$ & $5.85$ & $9.1$ & $37113$ & $141268$ & $-5.2$ & $0.25$ & $0.73$ & $-0.172$ & $-6.3$\\
$19$ & $7$ & $33$ & $0.0161$ & $5.91$ & $3.17$ & $63351$ & $144980$ & $-4$ & $0.15$ & $0.84$ & $-0.873$ & $-6.29$\\
$19$ & $7$ & $110$ & $0.0191$ & $5.94$ & $3.61$ & $59453$ & $145901$ & $-4.1$ & $0.16$ & $0.82$ & $-0.933$ & $-6.28$\\
$19$ & $7$ & $201$ & $0.0192$ & $5.95$ & $3.63$ & $59294$ & $146165$ & $-4.2$ & $0.16$ & $0.82$ & $-0.956$ & $-6.28$\\
$19$ & $7$ & $367$ & $0.0202$ & $5.97$ & $3.69$ & $58867$ & $146630$ & $-4.2$ & $0.16$ & $0.82$ & $-0.97$ & $-6.28$\\
$19$ & $7$ & $669$ & $0.0211$ & $5.99$ & $3.79$ & $58120$ & $147110$ & $-4.2$ & $0.16$ & $0.82$ & $-0.979$ & $-6.28$\\
$22$ & $8$ & $5$ & $0.005$ & $6.7$ & $1.68$ & $91358$ & $177100$ & $-3.4$ & $0.06$ & $0.92$ & $-0.807$ & $-6.17$\\
$25$ & $9$ & $5$ & $0.005$ & $7.99$ & $1.43$ & $105960$ & $230583$ & $-3.3$ & $0.06$ & $0.92$ & $-0.842$ & $-6.01$\\
$25$ & $9$ & $10$ & $0.0249$ & $7.7$ & $17.34$ & $29929$ & $216801$ & $-6.2$ & $0.38$ & $0.6$ & $0.05$ & $-5.06$\\
\hline
\hline
\end{tabular}
\footnotesize
\label{Table:Models5}
\end{threeparttable}
\label{tab:modelsV17c}
\end{table*}
\begin{table*}
\centering
\caption{Initial parameters and final properties for stellar evolution calculations with $0.45>q>0.35$ with the NL00 prescription.}
\begin{threeparttable}
\begin{tabular}{ccc|cccccccccc}
\hline
$M_1/$ & $M_2/$ & $P_\mathrm{i}/$ & $M_\mathrm{H}/$ & $M/$ & $R/$ & $T_\mathrm{eff}/$ & $L/$ & $M_v$ & $X_\mathrm{s}$ & $Y_\mathrm{s}$ & $f_\mathrm{L}$ &$\log_{10} \left( | \dot{M} | /  \mathrm{M}_\odot \, \mathrm{yr}^{-1}\right)$\\
$\mathrm{M}_\odot$ & $\mathrm{M}_\odot$ & $\mathrm{d}$ & $\mathrm{M}_\odot$ & $\mathrm{M}_\odot$ & $\mathrm{R}_\odot$ & $\mathrm{K}$ & $\mathrm{L}_\odot$ & & & & & \\
\hline
$12$ & $5$ & $10$ & $0$ & $2.75$ & $9.5$ & $26980$ & $42948$ & $-4.6$ & $0$ & $0.98$ & $0.017$ & $-5.9$\\
$12$ & $5$ & $18$ & $0$ & $2.79$ & $8.82$ & $28148$ & $43849$ & $-4.6$ & $0$ & $0.98$ & $-0.215$ & $-5.89$\\
$12$ & $5$ & $33$ & $0$ & $2.81$ & $8.32$ & $29104$ & $44605$ & $-4.5$ & $0$ & $0.98$ & $-0.536$ & $-5.88$\\
$12$ & $5$ & $60$ & $0$ & $2.83$ & $7.77$ & $30217$ & $45168$ & $-4.4$ & $0$ & $0.98$ & $-0.744$ & $-5.88$\\
$12$ & $5$ & $110$ & $0$ & $2.85$ & $7.54$ & $30752$ & $45650$ & $-4.4$ & $0$ & $0.98$ & $-0.843$ & $-5.87$\\
$12$ & $5$ & $201$ & $0$ & $2.87$ & $7.42$ & $31060$ & $46092$ & $-4.4$ & $0$ & $0.98$ & $-0.902$ & $-5.86$\\
$12$ & $5$ & $367$ & $0$ & $2.89$ & $7.08$ & $31895$ & $46601$ & $-4.3$ & $0$ & $0.98$ & $-0.939$ & $-5.86$\\
$12$ & $5$ & $669$ & $0$ & $2.92$ & $6.62$ & $33173$ & $47653$ & $-4.3$ & $0$ & $0.98$ & $-0.964$ & $-5.85$\\
$14$ & $5$ & $5$ & $0$ & $3.31$ & $4.49$ & $42672$ & $60004$ & $-3.9$ & $0$ & $0.98$ & $-0.184$ & $-5.72$\\
$14$ & $5$ & $10$ & $0$ & $3.38$ & $4.16$ & $44796$ & $62677$ & $-3.9$ & $0$ & $0.98$ & $-0.509$ & $-5.69$\\
$14$ & $5$ & $18$ & $0$ & $3.44$ & $3.97$ & $46154$ & $64397$ & $-3.8$ & $0$ & $0.98$ & $-0.69$ & $-5.68$\\
$14$ & $5$ & $33$ & $0$ & $3.48$ & $3.82$ & $47306$ & $65819$ & $-3.8$ & $0$ & $0.98$ & $-0.802$ & $-5.67$\\
$14$ & $5$ & $60$ & $0$ & $3.51$ & $3.72$ & $48180$ & $66928$ & $-3.8$ & $0$ & $0.98$ & $-0.871$ & $-5.66$\\
$14$ & $5$ & $110$ & $0$ & $3.54$ & $3.63$ & $48939$ & $67886$ & $-3.8$ & $0$ & $0.98$ & $-0.915$ & $-5.65$\\
$14$ & $5$ & $201$ & $0$ & $3.56$ & $3.57$ & $49479$ & $68571$ & $-3.7$ & $0$ & $0.98$ & $-0.943$ & $-5.64$\\
$14$ & $5$ & $367$ & $0$ & $3.58$ & $3.51$ & $50005$ & $69229$ & $-3.7$ & $0$ & $0.98$ & $-0.963$ & $-5.64$\\
$16$ & $6$ & $5$ & $0$ & $3.98$ & $2.5$ & $61895$ & $82629$ & $-3.4$ & $0$ & $0.98$ & $-0.636$ & $-5.54$\\
$16$ & $6$ & $10$ & $0$ & $4.06$ & $2.35$ & $64433$ & $85341$ & $-3.4$ & $0$ & $0.98$ & $-0.786$ & $-5.52$\\
$16$ & $6$ & $18$ & $0$ & $4.12$ & $2.26$ & $65965$ & $87024$ & $-3.3$ & $0$ & $0.98$ & $-0.862$ & $-5.51$\\
$16$ & $6$ & $33$ & $0$ & $4.16$ & $2.19$ & $67284$ & $88477$ & $-3.3$ & $0$ & $0.98$ & $-0.911$ & $-5.5$\\
$16$ & $6$ & $60$ & $0$ & $4.2$ & $2.14$ & $68325$ & $89671$ & $-3.3$ & $0$ & $0.98$ & $-0.941$ & $-5.49$\\
$16$ & $6$ & $110$ & $0$ & $4.23$ & $2.09$ & $69301$ & $90788$ & $-3.3$ & $0$ & $0.98$ & $-0.962$ & $-5.49$\\
$16$ & $6$ & $201$ & $0$ & $4.26$ & $2.06$ & $70045$ & $91679$ & $-3.3$ & $0$ & $0.98$ & $-0.975$ & $-5.48$\\
$16$ & $6$ & $367$ & $0$ & $4.28$ & $2.03$ & $70782$ & $92554$ & $-3.3$ & $0$ & $0.98$ & $-0.983$ & $-5.47$\\
$19$ & $7$ & $5$ & $0$ & $4.64$ & $1.56$ & $83159$ & $104662$ & $-3$ & $0$ & $0.98$ & $-0.822$ & $-5.41$\\
$19$ & $7$ & $10$ & $0$ & $4.79$ & $1.47$ & $86849$ & $110022$ & $-3$ & $0$ & $0.98$ & $-0.873$ & $-5.38$\\
$19$ & $7$ & $33$ & $0$ & $5.55$ & $1.4$ & $93983$ & $137913$ & $-3$ & $0$ & $0.98$ & $-0.941$ & $-5.25$\\
$19$ & $7$ & $110$ & $0$ & $5.54$ & $1.39$ & $94400$ & $137558$ & $-3$ & $0$ & $0.98$ & $-0.976$ & $-5.25$\\
$19$ & $7$ & $201$ & $0$ & $5.55$ & $1.39$ & $94472$ & $138110$ & $-3$ & $0$ & $0.98$ & $-0.984$ & $-5.25$\\
$19$ & $7$ & $367$ & $0$ & $5.74$ & $1.44$ & $93589$ & $143815$ & $-3.1$ & $0$ & $0.98$ & $-0.988$ & $-5.23$\\
$19$ & $7$ & $669$ & $0.0098$ & $5.9$ & $2.75$ & $68166$ & $146545$ & $-3.8$ & $0.15$ & $0.83$ & $-0.985$ & $-5.34$\\
$22$ & $8$ & $5$ & $0$ & $5.17$ & $1.18$ & $99642$ & $123196$ & $-2.8$ & $0$ & $0.98$ & $-0.868$ & $-5.31$\\
$22$ & $8$ & $10$ & $0$ & $5.57$ & $1.22$ & $100897$ & $138060$ & $-2.9$ & $0$ & $0.98$ & $-0.903$ & $-5.25$\\
$25$ & $9$ & $5$ & $0$ & $5.94$ & $0.96$ & $115772$ & $147516$ & $-2.6$ & $0$ & $0.98$ & $-0.9$ & $-5.21$\\
$25$ & $9$ & $10$ & $0.0248$ & $7.7$ & $17.22$ & $30035$ & $216838$ & $-6.2$ & $0.38$ & $0.6$ & $0.044$ & $-5.02$\\
\hline
\hline
\end{tabular}
\footnotesize
\label{Table:Models6}
\end{threeparttable}
\label{tab:modelsNL00c}
\end{table*}

The evolution of the primary as a single star after core carbon depletion was continued until iron core collapse for 95 models. The remaining time until core collapse was found to be $\Delta t \la 30 \, \mathrm{yr}$. For models with $M \la 7 \, \mathrm{M}_\odot$ at the end of the binary evolution the remaining time closely follows the relation
\begin{eqnarray}
\begin{aligned}
\log_{10} \left( \Delta t / \mathrm{yr} \right) \simeq 2.4985 - 1.8934 \log_{10} \left( M / \mathrm{M}_\odot \right),
\label{eq:dt}
\end{aligned}
\end{eqnarray}
where the mass $M$ at core carbon depletion is approximately the helium core mass\footnote{The helium core mass is tightly correlated to the carbon-oxygen core mass.}, because the hydrogen envelope is either of very low mass or non-existent. Models with $M \ga 7 \, \mathrm{M}_\odot$ do not follow equation (\ref{eq:dt}) and have much shorter time scales, with $\Delta t < 1 \, \mathrm{yr}$ for the highest masses. Models with $M \la 3.3 \, \mathrm{M}_\odot$ did not reach core collapse so we extrapolate with equation (\ref{eq:dt}) for the model with the lowest mass to a remaining time of $\Delta t \approx 48 \, \mathrm{yr}$. The range we find for $\Delta t$ is similar to the time scales for neon and oxygen burning in the core given by table 1 of \cite{Woosleyetal2002} with an additional delay time of several years between core carbon depletion and neon ignition. This shows that we can expect negligible changes between the end of our binary simulations and terminal iron core collapse.

\label{lastpage}

\end{document}